%% file: main.tex
\documentclass[twocolumn,aps,prl,superscriptaddress]{revtex4-1}
\usepackage{amsmath,amssymb}
\usepackage{graphicx}
\usepackage{xcolor}
\usepackage[colorlinks,bookmarks=false,citecolor=blue,linkcolor=blue,urlcolor=blue,hyperfootnotes=true]{hyperref}
\makeatother

\begin{document}
\title{Generating superconducting vortices via spin-orbit coupling}

\author{L. A. B. Olde Olthof} 
\email[]{labo2@cam.ac.uk}
\affiliation{Department of Materials Science \& Metallurgy, University of Cambridge, CB3 0FS Cambridge, United Kingdom}
\author{X. Montiel}
\email[]{x.montiel1@gmail.com}
\affiliation{Department of Materials Science \& Metallurgy, University of Cambridge, CB3 0FS Cambridge, United Kingdom}
\author{J. W. A. Robinson}
\email[]{jjr33@cam.ac.uk}
\affiliation{Department of Materials Science \& Metallurgy, University of Cambridge, CB3 0FS Cambridge, United Kingdom}
\author{A. I. Buzdin}
\email[]{alexandre.bouzdine@u-bordeaux.fr}
\affiliation{University Bordeaux, LOMA UMR-CNRS 5798, F-33405 Talence Cedex, France}
\affiliation{Department of Materials Science \& Metallurgy, University of Cambridge, CB3 0FS Cambridge, United Kingdom}

\date{\today}

\begin{abstract}
Spin-orbit coupling (SOC) is known to play an important role in superconductor/ferromagnet heterostructures. Here we demonstrate that SOC results in the spontaneous generation of vortices in an \textit{s}-wave superconductor placed below a ferromagnetic metal with intrinsic or interfacial SOC. By using the generalized London and Ginzburg-Landau theories, we calculate the supercurrent amplitude in various types of S/F heterostructures at and below the superconducting critical temperature. In particular, we show that even in the limit of weak SOC, vortex-antivortex pairs are stabilized and that an attractive interaction between vortices and SOC act to pin vortices along the S/F interface. We discuss key experiments to investigate these phenomena, which would provide a platform to the creation of Abrikosov vortex memory.

\end{abstract}
\maketitle
\noindent
In an \textit{s}-wave superconductor (S), charge currents flows in the absence of dissipation but since the Cooper pairs consist of electrons with antiparallel spins in a spin-singlet state, the current does not carry net spin. The opposite occurs in a ferromagnetic (F) metal in which an internal exchange field $h_\text{ex}$ creates a mismatch in the density of states of electrons with up and down spins, meaning that charge flow with a net spin-polarization but with with dissipitation \cite{BuzdinReview}. At a S/F interface, $h_\text{ex}$ interacts within S over the superconducting coherence length which for Nb is $\xi_S=30-40$ nm \cite{kittel}, whilst in F the interaction is much shorter with $\xi_F=1-3$ nm in Co \cite{length1,length2,length3}, Fe \cite{length2,length4}, Ni \cite{length1,length2,length3} due to $h_\text{ex}$ acting to rapidly dephase the singlet pairs. Consequently, there exists a F-thickness-dependent modulation of the critical temperature $T_c$ \cite{Radovic} in S/F bilayers and critical current $I_c$ oscillations in S/F/S Josephson junctions \cite{Bulaevskii,Buzdin81,Buzdin91}. In F$_1$/S/F$_2$ \cite{FSF1,FSF2,GennesFSF,Tagirov,BuzdinEPL} or F$_1$/F$_2$/S \cite{FFS1,FFS2,FFS3,FFS4} spin-valves, the total exchange field acting on the superconductivity is controllable via magnetization-alignment of the F layers, which creates a dependence of $T_c$ on magnetic configuration or $I_c$ in S/F$_1$/F$_2$/S Josephson junctions \cite{Bell,Nied,Samok}. Furthermore, a non-parallel alignment of magnetizations between the F layers converts singlet pairs to a spin-aligned triplet state \cite{Bergeret13,Bergeret14,Bobkova}. Since spin-aligned triplet pairs are spin-polarized, triplet supercurrents are able to carry a net spin in addition to charge and so are stable in F materials over length scales that exceed $\xi_F$ \cite{Keizer2006,Anwar2010,Khaire,triplet1,triplet2,triplet3,triplet4}. These phenomena form the basis of applications in superconducting spintronics \cite{Linder}.

Spin-orbit coupling (SOC) is also present at a S/F interface due to the absence of inversion symmetry \cite{Edelstein,Buzdin08,Bergeret13,Bergeret14,Mironov15,Jacobsen2} and can be enhanced by introducing an interlayer between the S and F layers with strong SOC. Since spin is not conserved in the presence of SOC, the SOC locally disrupts the singlet pairs, rotating their spins towards the direction of magnetization. Hence, SOC in combination with a single F layer can generate spin-aligned triplet pairs \cite{Takada,Gorkov}. For example, in a recent ferromagnetic resonance spin-pumping experiment on Pt/Nb/Py  \cite{Montiel,Kunrok3,Kunrok} spin-pumping efficiency was enhanced over the normal state below $T_c$. These results indicated the presence of a spin-triplet channel in Nb that exists over $\xi_S$, which forms due to SOC in Pt in combination with $h_\text{ex}$ from Py.     

Interfacial SOC provides a Rashba-type contribution $\sim (\vec{\sigma}\times \vec{p})\cdot \hat{n}$ to the energy, where $\vec{\sigma}$ is the electron spin, $\vec{p}$ is the orientation of the momentum and $\hat{n}$ is the unit vector along the broken inversion symmetry axis.
Magnetic order spin-polarizes electrons, meaning that momentum along $\vec{\sigma}\times\hat{n}$ is energetically favourable \cite{Mironov} and so may induce spontaneous charge currents. This is predicted near a S/F interface within the London penetration depth from the interface \cite{Mironov}, around magnetic islands on a thin-film S \cite{Pers}, in a closed S loop partially covered by a ferromagnetic insulator \cite{Robinson} and in a thin-film S in contact with a Ne\'el skyrmion \cite{Baumard}. 

In this \textit{Letter} we investigate the interaction of SOC with vortices in S/F heterostructures, at temperatures $T$ below and at $T_c$. We derive a criterion for spontaneous vortex generation due to SOC and show that this is achievable using realistic materials parameters. Based on these results, we demonstrate that SOC strongly affects transport properties in S/F junctions.

\begin{figure}[tb]
    \centering
    \includegraphics[width=\linewidth]{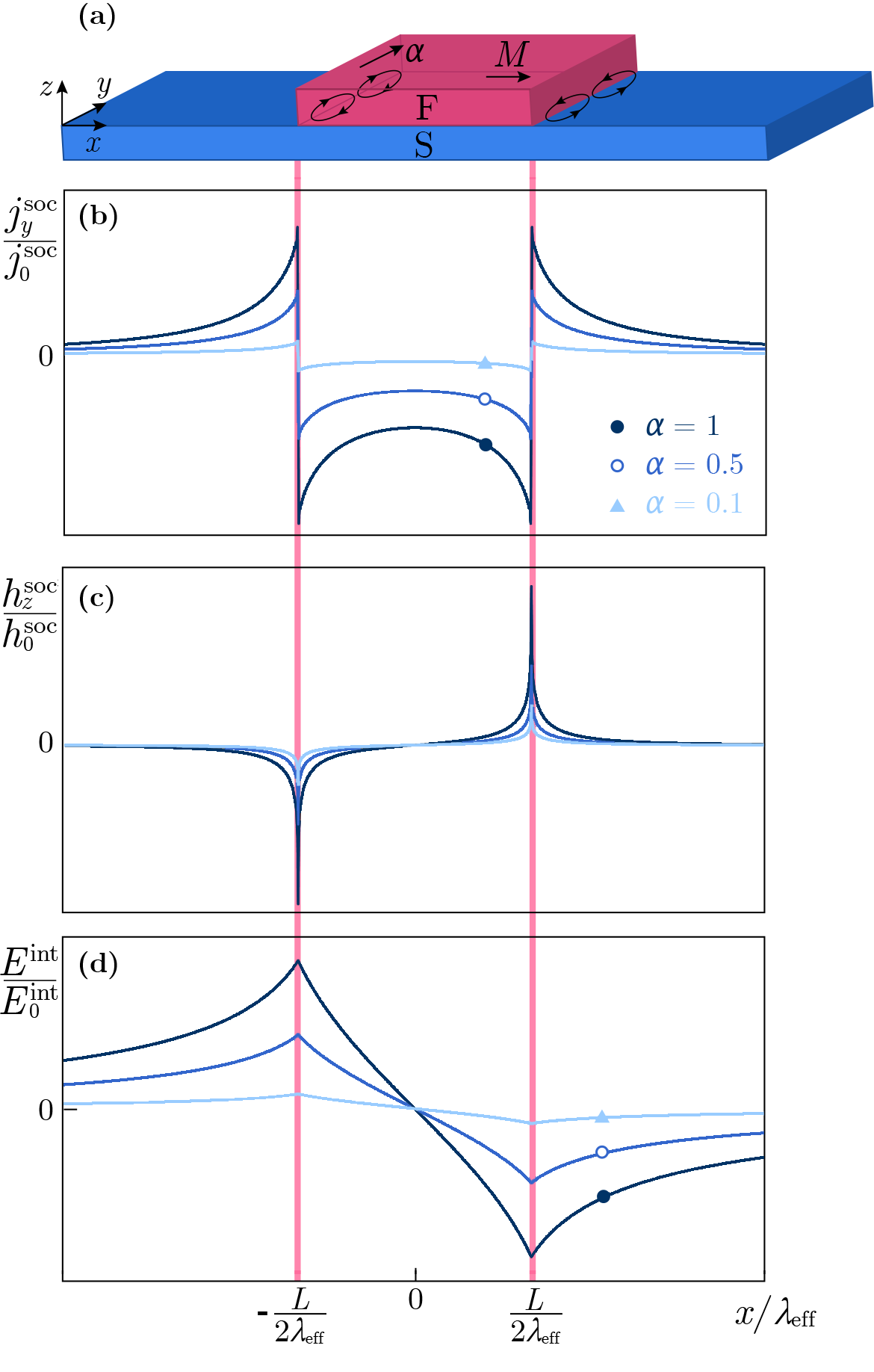}
    \caption{A ferromagnetic strip on a superconductor at $T\ll T_c$. {\bf (a)} A schematic illustration of S/F heterostructure with magnetization $\vec{M}$ and SOC $\vec{\alpha}$. Vortices and antivortices are represented by circular currents at the interfaces. {\bf (b)} The current $j_y^\text{soc}$ and {\bf (c)} the magnetic field $h_z^\text{soc}$ due to SOC. {\bf (d)} The interaction energy between the vortices and SOC $E^\text{int}$.}
    \label{Fig1}
\end{figure}

We model a type-II S of thickness $d_S$ (with $d_S<\xi$) that is partially covered by a thin-film metallic F with thickness $d_F$, as schematically illustrated in Fig.~\ref{Fig1}(a). We use the general Ginzburg-Landau (GL) approach to describe the S/F system where the density of the free energy $F=\int f(\vec{r})d^3\vec{r}$ is 
\begin{align}
    f =&\ a|\psi|^2+\frac{b}{2}|\psi|^4 + \frac{1}{4m} |\hat{D} \psi |^2 \nonumber \\ 
    &+ \frac{h^2}{8\pi} + \frac{\vec{\alpha}}{4m} \cdot\left(\psi^*\hat{D}\psi + \psi(\hat{D}\psi)^*\right), \label{f}
\end{align}
where $a = a_0 (T-T_c)/T_c$ and $b$ are the standard GL coefficients, $T$ is temperature, $\psi$ is the superconducting order parameter, $\hat{D}=-i\vec{\nabla} + 2e\vec{A}$ is the gauge-invariant momentum operator with $\hbar=c=1$ and $\vec{h}$ is the magnetic induction.
We assume in-plane magnetization in F with the magnetization pointing along $x$ resulting in SOC along $y$.
SOC in the F is described by $\vec{\alpha} = \alpha \ \Theta\left(x-L/2\right)\Theta\left(L/2-x\right) \ \hat{y}$, where $\Theta(x)$ is the Heaviside step function and the SOC strength $\alpha$ is \cite{Baumard}
\begin{equation}
    \alpha \simeq \phi_0 \frac{v_R}{v_F} \frac{c}{d_S}\frac{h_\text{ex}}{T_c\xi}, \label{alpha}
\end{equation}
where $\phi_0/2\pi =1/2e$ is the flux quantum, $v_R$ and $v_F$ are the Rashba and Fermi velocities, respectively, and $c$ is the atomic lattice parameter.
The gradient over the order parameter appears only in the presence of both SOC and a ferromagnetic exchange field \cite{Mineev}.

For $T\ll T_c$, superconductivity is well-developed, such that the GL order parameter is $\psi = |\psi|e^{i\varphi}$, where $|\psi|$ is constant and $\varphi$ is the phase of the superconducting order parameter \cite{supp}. Following Ref.~\cite{Degennes}, we introduce the vorticity
\begin{equation}
    \vec{\Phi}=\phi_0\frac{\hat{n}\times(\vec{r}-\vec{r}_0)}{(r-r_0)^2} \label{vortex}
\end{equation}
to describe the phase singularity related to a vortex present at $\vec{r}=\vec{r}_0$.
Assuming $d_S\ll\lambda$, superconductivity along the $z$-axis can be taken constant, such that its integral yields a factor $d_S$. We define the effective penetration depth as $\lambda_\text{eff} = \lambda^2/d_S$ \cite{Degennes}.
The GL free energy is then
\begin{equation}
    F=F_0 + \frac{1}{8\pi\lambda_\text{eff}} \int \left(\vec{\Phi} - \vec{A} + \frac{\vec{\alpha}\phi_0}{2\pi} \right)^2 d^2\vec{r} + \int \frac{h^2}{8\pi} d^3\vec{r},\label{Fgl}
\end{equation}
where $F_0$ is the free energy in the normal state, above $T_c$. 

The current density in the plane $z=0$ is given by $\vec{j} = -\partial f/\partial\vec{A}\ \delta(z)$.
From the Maxwell-Ampere equation $4\pi \vec{j} = \vec{\nabla}\times\vec{h} = -\Delta\vec{A}$, we obtain
\begin{equation}
    -\Delta\vec{A} = \frac{1}{\lambda_\text{eff}}\left( \vec{\Phi} -\vec{A} + \frac{\vec{\alpha}\phi_0}{2\pi} \right)\delta(z),  \label{deA}
\end{equation}
which in Fourier space, has solution $\vec{A}_q$ (the two-dimensional Fourier transform of $\vec{A}(z=0)$)
\begin{equation}
 \vec{A}_q = \frac{\vec{\Phi}_q +\vec{\alpha}_q\phi_0/2\pi}{1 + 2|\vec{q}|\lambda_\text{eff}}.\label{Aq}
\end{equation}
$\vec{A}_q$ consists of two contributions $\vec{\Phi}_q$ and $\vec{\alpha}_q$, which are terms that describe the two-dimensional Fourier transforms of the vortices $\vec{\Phi}$ and SOC $\vec{\alpha}$, respectively \cite{supp}.
Consequently, the current density also contains two terms, i.e. $\vec{j}= \vec{j}^\text{vor}+\vec{j}^\text{soc}$. The vortex contribution $\vec{j}^\text{vor}$ is given in Ref.~\cite{Degennes}. The current generated due to SOC is
%\small
%\begin{equation}
%    \vec{j}^\text{soc}(x) = \frac{\alpha\phi_0\hat{y}}{4\pi^3} \int_0^\infty %\frac{\sin\left(q_x(\frac{L}{2}-x)\right)+\sin\left(q_x(\frac{L}{2}+x)\right)}{1 + %2q_x\lambda_\text{eff}} dq_x. \label{jso}
%\end{equation}
%\normalsize
\small
\begin{equation}
    \vec{j}^\text{soc}(x) = j_0^{soc}\hat{y} \int_0^\infty \frac{\sin\left(\frac{p(\frac{L}{2}-x)}{\lambda_\text{eff}}\right)+\sin\left(\frac{p(\frac{L}{2}+x)}{\lambda_\text{eff}}\right)}{1 + 2p} dp \label{jso}
\end{equation}
\normalsize
with $j_0^\text{soc}=\frac{\alpha \phi_0}{4\pi^3 \lambda_\text{eff}}$ and $p=q_x\lambda_\text{eff}$. 
The local magnetic field at the surface of the superconductor is obtained by transforming the Maxwell equation $\vec{h} = \vec{\nabla}\times\vec{A}$ into Fourier space. The magnetic field consists of two contributions $\vec{h}= \vec{h}^\text{vor}+\vec{h}^\text{soc}$, of which the first is known \cite{Degennes} and the SOC contribution is
%\small
%\begin{equation}
%    \vec{h}^\text{soc}(x)
%    = \frac{\alpha\phi_0 \hat{z}}{2\pi^2}\int_0^\infty\frac{\cos\left(q_x(\frac{L}{2}-x)\right)-\cos\left%(q_x(\frac{L}{2}+x)\right)}{1 + 2q_x\lambda_\text{eff}}dq_x.  \label{hso}
%\end{equation}
%\normalsize
\small
\begin{equation}
    \vec{h}^\text{soc}(x)
    = h_0^{soc}\hat{z} \int_0^\infty \frac{\cos\left(\frac{p(\frac{L}{2}-x)}{\lambda_\text{eff}}\right)+\cos\left(\frac{p(\frac{L}{2}+x)}{\lambda_\text{eff}}\right)}{1 + 2p} dp  \label{hso}
\end{equation}
\normalsize
with $h_0^{soc}=\frac{\alpha \phi_0}{2\pi^2 \lambda_\text{eff}}$.
The interaction between the vortices and SOC is described by the cross terms $(\sim \vec{\alpha}_q\cdot \vec{\Phi}^*_q+ \vec{\alpha}^*_q\cdot\vec{\Phi}_q)$ in the total free energy~(\ref{Fgl}). We obtain
%\begin{equation}
%    E^\text{int}(x) = -\frac{\alpha\phi_0^2}{2\pi^3}\int_0^\infty %\frac{\sin\left(\tfrac{1}{2}q_xL\right)\sin\left(q_x x\right)}{q_x(1+2q_x\lambda_\text{eff})} dq_x. %\label{Eint}
%\end{equation}
\begin{equation}
    E^\text{int}(x) = -E_0^\text{int}\int_0^\infty \frac{\sin\left(\frac{pL}{2\lambda_\text{eff}}\right)\sin\left(\frac{px}{\lambda_\text{eff}}\right)}{p(1+2p)} dp \label{Eint}
\end{equation}
with $E_0^\text{int}=\frac{\alpha\phi_0^2}{2\pi^3}$.

The current density~(\ref{jso}), magnetic field~(\ref{hso}) and interaction energy~(\ref{Eint}) are displayed in Fig.~\ref{Fig1}(b)-(d). Since the SOC is an interface effect located inside the F, the current goes down towards the edges of the F and has sharp peaks just outside of it. The SOC causes local magnetic field spikes just at the interfaces, which are positive and negative, corresponding to the magnetization direction. The interaction energy attracts vortices towards its minimum at $-L/2$, while it repels vortices from its maximum at $L/2$ (an antivortex).
All features become more prominent with increasing SOC strength $\alpha$. 

The behaviour of $\vec{j}^\text{soc}(x)$ and $E^\text{int}(x)$ follows the same trend in Ref.~\cite{Vlasov}, even though the physical origin in Ref.~\cite{Vlasov} is electromagnetic, and not SOC. 
The interaction between the vortex and edge of the F considered in \cite{Vlasov} is generated by stray fields, which is proportional to the magnetization $M$ and $d_F$. The SOC mechanism considered here dominates if $\alpha\phi_0/2\pi^2 > 2Md_F$. Taking into account that the lower critical field of the S is $H_{c1} = (\phi_0/4\pi\lambda^2) \ln(\lambda/\xi)$, this condition can be written as $\alpha>B(0)\pi d_F/H_{c1}\lambda^2$, where $B(0)=4\pi M$ is the magnetic induction in the F. For $B(0)\sim H_{c1}$ and $d_f\sim d_S \sim\xi$, this condition is always fulfilled for typical values of $\alpha$ needed for vortex generation (see Eq.~(\ref{cond}) below). The sign of the current generated by the SOC mechanism may be opposite to the current due to the electromagnetic interaction, because it depends on the sign of the exchange integral determining $h_\text{ex}$.

We use Eq.~(\ref{Eint}) to derive the criterion for energetically favourable vortex and antivortex nucleation. The total free energy of the S/F bilayer consists of energy originating from the SOC, the vortices and the interaction between them. The first term is independent of vortices and hence vortices appear when the negative interaction energy is balanced with the positive energy from the vortices - i.e. $E^\text{vor} + E^\text{int} < 0$. The energy of vortices $E^\text{vor}$ is \cite{Degennes}
\begin{equation}
    E^\text{vor} = \left(\frac{\phi_0}{4\pi}\right)^2\frac{1}{\lambda_\text{eff}} \ln\left(\frac{\lambda_\text{eff}}{\xi}\right).
\end{equation}
In the case of $L\gg \lambda_\text{eff}$, the interaction energy between a vortex at $x=L/2$ and the SOC is 
\begin{equation}
  E^\text{int} \simeq -\frac{\alpha\phi_0^2}{4\pi^3}\ln\left(\frac{L}{\lambda_\text{eff}}\right). \label{Eapprox}
\end{equation}
Comparing $E^\text{vor}$ and $E^\text{int}$, we obtain the SOC strength required for the formation of a vortex-antivortex pair in the absence of an external magnetic field
\begin{align}
    \alpha > \frac{\pi}{4\lambda_\text{eff}}\frac{\ln\left(\dfrac{\lambda_\text{eff}}{\xi}\right)}{\ln\left(\dfrac{L}{\lambda_\text{eff}}\right)}, \label{cond}
\end{align}
where $\alpha$ is given by Eq.~(\ref{alpha}).
Qualitatively, this shows that an increase in $h_\text{ex}$ or $L$ or a reduction in $d_S$ facilitate vortex formation.
Quantitatively, typical ratios between the parameters are $v_R/v_F\sim 0.1$ and $ah_\text{ex}/d_ST_c \sim 0.1$ \cite{Robinson}, such that $\alpha \simeq 0.01/\xi$.
We assume $\xi/\lambda_\text{eff}\sim 10^{-3}$ and $\xi \ll L$. This implies that condition~(\ref{cond}) is easily fulfilled and the system will spontaneously generate vortex-antivortex pairs.
We may expect the antivortex chain to appear under the left edge of the F in Fig.~\ref{Fig1}(a), while a vortex chain appears under the right edge.
\begin{figure}[tb]
    \centering
    \includegraphics[width=.9\linewidth]{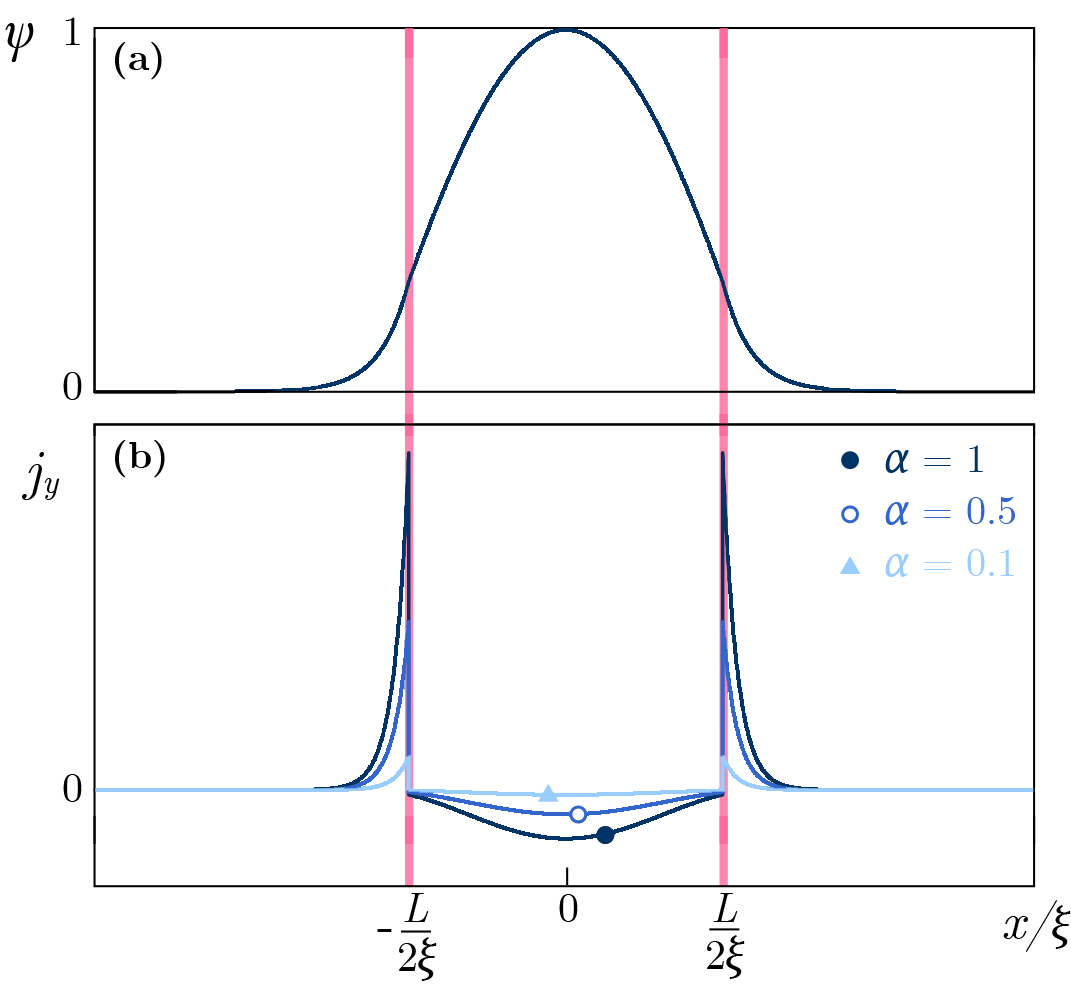}
    \caption{A ferromagnetic strip on a superconducting film at $T= T_c$. Spatial dependence of {\bf (a)} the normalized wave function $\psi(x)$ and {\bf (b)} the current density $j_y(x)$ for different (labelled) values of the SOC strength parameter $\alpha$.}
    \label{Fig2}
\end{figure}
\\\\
So far, we have considered the case where the superconductivity is well-developed. For $T\lesssim T_c$, the superconductivity is weak and we have to calculate the order parameter space dependency explicitly \cite{supp}.
Minimizing $f$ [Eq.~(\ref{f})], the GL equation in the absence of a magnetic field is
\begin{equation}
    a\psi + b |\psi|^2\psi - \frac{1}{4m}\left(\frac{\partial^2\psi}{\partial x^2}+\frac{\partial^2\psi}{\partial y^2}\right) - \frac{i \alpha}{2m} \frac{\partial\psi}{\partial y} = 0. \label{GL} 
\end{equation}
At $T_c$, the order parameter magnitude vanishes ($|\psi|^2\to 0$) and the GL equation can be linearized \cite{supp}. Solving Eq.~(\ref{GL}) for $\psi$ and directly calculating the current $\vec{j}$, we find \cite{supp}
\begin{align}
    \vec{j}(x) = \begin{cases}
    \dfrac{e\alpha}{m}|\psi(x)|^2 \hat{y} & \mbox{ for } |x|>L/2, \\\\
    -\dfrac{e\alpha\varepsilon}{m}|\psi(x)|^2 \hat{y} & \mbox{ for } |x|<L/2, \\
    \end{cases} \label{current}
\end{align}
where $\varepsilon$ is a small difference between the modulation vector and $\alpha$, that is found self-consistently \cite{supp}.

\begin{figure}[tb]
    \centering
    \includegraphics[width=\linewidth]{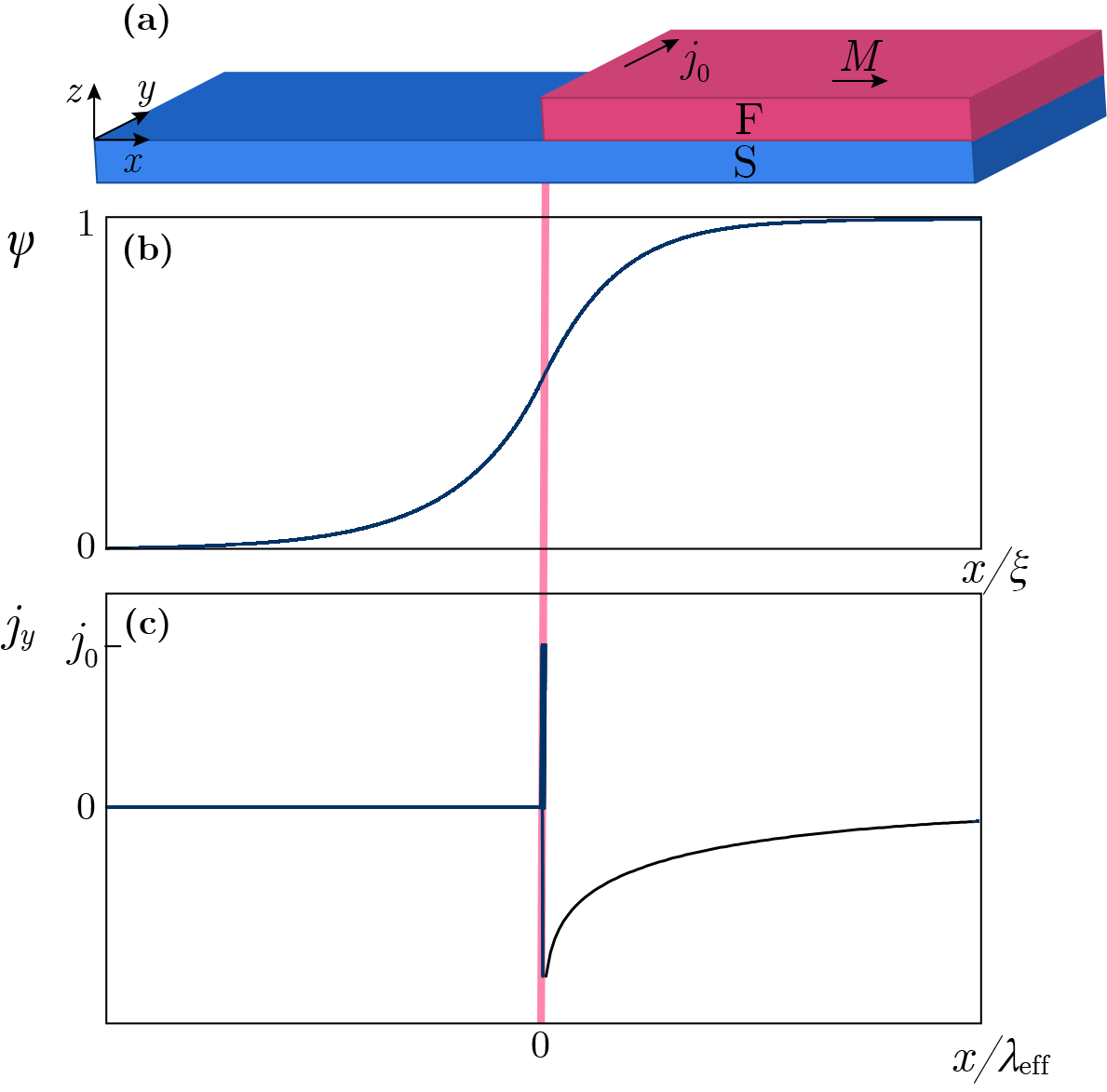}
    \caption{A superconducting film half covered by a ferromagnetic metal with $T\lesssim T_c$. {\bf (a)} A schematic illustration of S/F heterostructure with magnetization $\vec{M}$ and edge current $j_0\hat{y}$. Spatial dependence of {\bf (b)} the normalized wave function $\psi(x)$ and {\bf (c)} the current density $j_y(x)$.}
    \label{Fig3}
\end{figure}

The wave function $\psi(x)$ and current density $j_y(x)$ from Eq.~(\ref{current}) are shown in Fig.~\ref{Fig2}. The current density $j_y(x)$ exhibits large peaks at the interfaces and a small suppression inside the F. 
Comparing Fig.~\ref{Fig2}(b) to~\ref{Fig1}(b), the current goes up near the interface at $T=T_c$, which results from suppression of the order parameter.
\\\\
In general, when $T\lesssim T_c$, the order parameter can be written as $\psi(x,y) = \psi_0e^{iq_yy}$.
Substituting this into Eq.~(\ref{GL}), we find that $T-T_c$ has a maximum for $q_y=-\alpha$. Subsequently, we use $\psi(x,y) = \psi(x) e^{-i\alpha y}$, apply separation of variables to Eq.~(\ref{GL}) and evaluate the derivatives with respect to $y$ separately. These can be considered as a constant offset to $T_c$ that results in local modulations in $T_c$. In the presence of SOC, $T_c$ increases according to
\begin{align}
    & T_{c+} = T_c \left( 1 + \frac{\alpha^2}{4ma_0}\right). \label{Tc}
\end{align}
We note that $T_c$ in Eq.~(\ref{Tc}) relates to the critical temperature of the F/S bilayer without SOC, meaning that $T_{c+}-T_c$ represents a recovery of the critical temperature.
\\\\
 We use the modulated critical temperature~(\ref{Tc}) to consider an F covering the half space (i.e. $x\in[0,\infty)$, see Fig.~\ref{Fig3}(a)), with $T\lesssim T_c$.
We use Eq.~(\ref{Tc}) to solve Eq.~(\ref{GL}).

The total current consists of a current along the interface and the supercurrent - i.e.
\begin{equation}
    \vec{j}(x) = j_0\delta(x)\hat{y}-\frac{1}{4\pi\lambda_\text{eff}}\vec{A}(x). \label{j}
\end{equation}
Modelling the interface current as a Dirac $\delta$-function is validated by $\xi\ll\lambda$.
We use Biot-Savart's law $\vec{A} = \int \vec{j}/\sqrt{x^2+y^2} \ d\vec{l}$, which results in an implicit equation for $\vec{A}$. To avoid divergence issues, we first integrate over a strip of width $L$ and set $z=0$. We obtain
\begin{align}
    A_y(x) =&\ \ln|x-L|-\ln|x| \nonumber \\
    &+ \frac{1}{2\pi}\int_0^L A_y(x') \ln|x-x'| dx'. \label{Ay}
\end{align}

We solve this equation for $A_y(x)$ iteratively, let $L\to\infty$ and obtain the current density from Eq.~(\ref{j}).
The normalized wave function $\psi(x)$ and current density $j_y(x)$ from Eq.~(\ref{j}) are displayed in Fig.~\ref{Fig3}. The amplitude of the wave function increases near the interface and saturates inside F. The current is zero outside F and has a $\delta$-peak resulting from the edge current, then drops and gradually recovers inside F.
\\\\
In summary, we have theoretically investigated the role of SOC in S/F devices at and below $T_c$. We have shown that even weak SOC ($\alpha\simeq 0.01/\xi$) generates vortex-antivortex pairs and that an attractive interaction between vortices and SOC acts to pin vortices along S/F interfaces.
This effect would manifest experimentally as a peak in magnetic field profile across a S/F interface below $T_c$ and hence by measuring local magnetic fields through $T_c$ it should possible to distinguish the effect of SOC from stray fields from out-of-plane Bloch domain walls. Vortex generation and pinning are prominent when the magnetization is perpendicular to the interface and diminishes as the magnetization rotates in-plane towards the S/F interface. In our calculations, we consider a F metal with intrinsic SOC; however, matching behaviour is expected when separating the SOC from the F layer by introducing a metallic interface with intrinsic SOC \cite{Jacobsen}. Ideal materials combinations include Pt/Co, Pd/Co, Pt/Py and Pd/Py, which have high interfacial transparency with Nb \cite{Tao,Ghosh}. A key breakthrough of our theory is that it predicts the potential for vortex guiding without the usual requirement of stray magnetic fields to generate vortex-antivortex pairs and that vortex motion [see Fig.~\ref{Fig1}(a)] is reversible with applied current direction or magnetization which are key requirements for Abrikosov memory. 

\begin{acknowledgments}
The research was funded by EU COST CA16218  Nanocohybri (A.I.B.; J.W.A.R), the French ANR project SUPERTRONICS and OPTOFLUXONICS (A.I.B.), and the EPSRC (L.A.B.O.O., X.M. and J.W.A.R.) through an International Network and Programme Grants (EP/P026311/1; EP/N017242/1). 
\end{acknowledgments}

\bibliography{references}

\input{supp}

\end{document}

%% file: supp.tex
\pagebreak
%Latex settings for supplementary:
\setcounter{equation}{0}
\setcounter{figure}{0}
\setcounter{page}{1}
\makeatletter
\renewcommand{\theequation}{S\arabic{equation}}
\renewcommand{\thefigure}{S\arabic{figure}}
\renewcommand{\bibnumfmt}[1]{[S#1]}
\renewcommand{\citenumfont}[1]{S#1}

\clearpage
\onecolumngrid
\begin{center}
\textbf{\large Supplemental Materials: \\  Generating a superconducting vortex using spin-orbit coupling}\\
\vspace{.5cm}
L. A. B. Olde Olthof,$^1$ X. Montiel,$^1$ J. W. A. Robinson,$^1$ and A. I. Buzdin$^{2,1}$\\
\small{ \it{$^1$Department of Materials Science \& Metallurgy, University of Cambridge, CB3 0FS Cambridge, United Kingdom \\ $^2$University Bordeaux, LOMA UMR-CNRS 5798, F-33405 Talence Cedex, France}}
\end{center}
\section{Ferromagnetic strip}   
\twocolumngrid

\subsection{Free energy $F$}

As described in the main text, the free energy $F$ relates to the free energy density as $F=\int f(\vec{r})d^3\vec{r}$ and consists of several terms
\begin{align}
   F =&\int \left(\ a|\psi|^2+\frac{b}{2}|\psi|^4 + \frac{h^2}{8\pi}\right) d^3\vec{r}+ F_\text{s} + F_\text{soc}, \label{ff}
\end{align}
where $F_\text{s}$ is the kinetic energy term related to superconductivity and $F_\text{soc}$ is the free energy produced by the SOC.

The term $F_\text{s}$ relates to the third term in Eq.~(1) of the main text as $F_\text{s}= \int \frac{1}{4m} |\hat{D} \psi |^2\ d^3\vec{r} $ . The wave function is $\psi = |\psi|e^{i\varphi}$, such that $\psi^*\vec{\nabla}\psi-\psi\vec{\nabla}\psi^*=2i|\psi|^2\vec{\nabla}\varphi$. This relation holds in the regime $T \ll T_c$, where the superconductivity can be considered uniform ($|\psi|$ is a constant). The term can then be expanded as
\begin{align}
    &\left| \left(-\frac{i}{2e}\vec{\nabla} + \vec{A} \right) \psi \right|^2 \nonumber \\
    &= \frac{1}{4e^2}\vec{\nabla}\psi\vec{\nabla}\psi^*- \frac{i}{2e}(\psi^*\nabla\psi-\psi\nabla\psi^*)\vec{A} + \vec{A}^2\psi\psi^* \nonumber \\
    &= \left(\frac{1}{4e^2}(\vec{\nabla}\varphi)^2 + \frac{1}{e}\vec{A}\vec{\nabla}\varphi + \vec{A}^2\right) |\psi|^2 \nonumber \\
    &= \left( \vec{\Phi}-\vec{A}\right)^2|\psi|^2, \label{rewriteD}
\end{align}
where $\vec{\Phi}$ is given by Eq.~(3) in the main text. The London penetration depth is defined as
\begin{equation}
    \lambda^2 = \frac{m}{8\pi e^2|\psi|^2}. \label{lambda}
\end{equation}
Substituting Eqs.~(\ref{rewriteD}) and (\ref{lambda}) in the third term of Eq.~(1) of the main text, the free energy from the superconductor becomes
\begin{equation}
    F_\text{s} = \frac{1}{8\pi\lambda^2} \int \left(\vec{\Phi} - \vec{A} \right)^2 d^3\vec{r}. \label{Fs}
\end{equation}
The term $F_\text{soc}$ relates with the last term in  Eq.~(1) of the main text as $F_\text{soc}=\int \frac{\vec{\alpha}}{4m} \cdot\left(\psi^*\hat{D}\psi + \psi(\hat{D}\psi)^*\right) d^3\vec{r}$. Using again Eqs.~(\ref{rewriteD}) and (\ref{lambda}), the free energy from the SOC becomes
\begin{equation}
     F_\text{soc} = \frac{1}{8\pi\lambda^2} \int \frac{\vec{\alpha}}{e}\cdot \left(\vec{\Phi}-\vec{A}\right) d^3\vec{r}. \label{Fsoc}
\end{equation}
Taking Eqs.~(\ref{Fs}) and (\ref{Fsoc}) together and using $1/2e=\phi_0/2\pi$ gives
\begin{align*}
    F_\text{s} + F_\text{soc} 
    &= \frac{1}{8\pi\lambda^2} \int  \left(\vec{\Phi}-\vec{A}+\frac{\vec{\alpha}\phi_0}{2\pi}\right)^2 - \frac{\vec{\alpha}^2\phi_0^2}{4\pi^2}\  d^3\vec{r}.
\end{align*}
The $\vec{\alpha}^2$ term is taken in the constant $F_0$ in Eq.~(4) of the main text.
The superconductivity is confined to a thin film with thickness $d_S$ (along the $z$-axis). We assume $d_S\ll \lambda$, such that the film is thin enough to take the integral constant along the $z$-axis and the integral over $z$ simply yields a factor $d_S$. We define the effective penetration depth $\lambda_\text{eff} = \lambda^2/d_S$. The $F_\text{s}+F_\text{soc}$ term reduces to
\begin{equation}
    F_\text{s}+F_\text{soc} = \frac{1}{8\pi\lambda_\text{eff}} \int \left(\vec{\Phi} - \vec{A} + \frac{\vec{\alpha}\phi_0}{2\pi} \right)^2 d^2\vec{r}, \label{ff4}
\end{equation}
which is the term in Eq.~(4) of the main text.

\subsection{Vector Potential $\vec{A}$}
The differential equation for $\vec{A}$ is 
\begin{equation}
    -\Delta\vec{A} = \frac{1}{\lambda_\text{eff}}\left( \vec{\Phi} -\vec{A} + \frac{\vec{\alpha}\phi_0}{2\pi} \right)\delta(z).  \label{deAs}
\end{equation}
To solve this equation for $\vec{A}$, we transform it to Fourier space with $\vec{q} = q_x\hat{x} + q_y\hat{y}$ and $\vec{k} \equiv k \hat{z}$. We introduce the following three and two-dimensional Fourier transforms:
\begin{align}
    & \vec{A}_{q,k} \equiv \int \vec{A}\ e^{i\vec{q}\cdot\vec{r}}e^{ikz} d^2\vec{r} \ dz, \nonumber\\
    & \vec{A}_q \equiv \frac{1}{2\pi}\int \vec{A}_{qk}\ dk = \int \vec{A}\ e^{i\vec{q}\cdot\vec{r}} d^2\vec{r}, \nonumber\\
    & \vec{\alpha}_q \equiv \int \vec{\alpha}\ e^{i\vec{q}\cdot\vec{r}} d^2\vec{r} = \frac{4\pi\alpha}{q_x}\sin\left(\frac{q_xL}{2}\right) \delta(q_y)\hat{y}, \label{fourieralpha} \\
    & \vec{\Phi}_q  \equiv \int \vec{\Phi}\ e^{i\vec{q}\cdot\vec{r}} d^2\vec{r} = i\phi_0 \frac{q_x \hat{y}-q_y\hat{x} }{|\vec{q}|^2}e^{iq_xL/2}. \label{fourierphi}
\end{align}
such that the Fourier transform of Eq.~(\ref{deAs}) becomes
\begin{equation}
    \vec{A}_{q,k} = \frac{1}{|\vec{q}|^2+k^2}\frac{1}{\lambda_\text{eff}}\left( \vec{\Phi}_q -\vec{A}_q +\frac{\vec{\alpha}_q\phi_0}{2\pi} \right). \label{Aqk}
\end{equation}
We perform integration over $k$, which yields the expression of the vector potential in Fourier space:
\begin{equation}
 \vec{A}_q = \frac{\vec{\Phi}_q +\vec{\alpha}_q\phi_0/2\pi}{1 + 2|\vec{q}|\lambda_\text{eff}}.\label{Aqs}
\end{equation}
The vector potential can be split into two contributions:
$\vec{A}_q = \vec{A}^\text{vor}_q + \vec{A}^\text{soc}_q $ with
\begin{align*}
&\vec{A}^\text{vor}_q=\frac{\vec{\Phi}_q}{1 + 2|\vec{q}|\lambda_\text{eff}}, &&\vec{A}^\text{soc}_q=\frac{\vec{\alpha}_q\phi_0/2\pi}{1 + 2|\vec{q}|\lambda_\text{eff}}.
\end{align*}.
\subsection{Current density $\vec{j}^\text{soc}(x)$}
From the Maxwell equation, we know $\vec{j} = -\Delta\vec{A}/4\pi$.
In Fourier space, this becomes
\begin{equation*}
    \vec{j}_q = \frac{|\vec{q}|}{2\pi} \vec{A}_q.
\end{equation*}
Since we are interested in the current originating from the SOC, we consider the vector potential $\vec{A}^\text{soc}_q$. Transforming it to real space, we obtain
\begin{align*}
    \vec{j}^\text{soc}(x) 
    &= \frac{\alpha\phi_0\hat{y}}{4\pi^3} \int \frac{\sin\left(\tfrac{1}{2}q_xL\right)}{1 + 2|\vec{q}|\lambda_\text{eff}} \frac{|\vec{q}|}{q_x} \delta(q_y) e^{-i\vec{q}\cdot\vec{r}} d^2\vec{q}.
\end{align*}
The integral over $q_y$ vanishes. To take the absolute value of $|\vec{q}|$ into account, we split the integral into $(-\infty,0)$ and $(0,\infty)$. Rewriting the result, we find
\small
\begin{equation}
    \vec{j}^\text{soc}(x) = \frac{\alpha\phi_0\hat{y}}{4\pi^3} \int_0^\infty \frac{\sin\left(q_x(\frac{L}{2}-x)\right)+\sin\left(q_x(\frac{L}{2}+x)\right)}{1 + 2q_x\lambda_\text{eff}} dq_x. \label{jsos}
\end{equation}
\normalsize
which is the relation (7) of the main text.

\subsection{Magnetic field $\vec{h}^\text{soc}(x)$}
The Maxwell equation $\vec{h} = \vec{\nabla}\times\vec{A}$ transformed into Fourier space becomes
\begin{equation}
    \vec{h}_{q,k} = i (\vec{q} + \vec{k})\times\vec{A}_{q,k}. \label{fourierh}
\end{equation}
We combine Eqs.~(\ref{Aqk}) and (\ref{Aqs}) and only consider the SOC term with $\vec{\alpha}_q$. We obtain
\begin{align}
    &\vec{h}^\text{soc}_{q,k} = \frac{2i|\vec{q}|}{(|\vec{q}|^2+k^2)(1 + 2|\vec{q}|\lambda_\text{eff})} (\vec{q} + \vec{k})\times \vec{\alpha}_q. \label{hqk}
\end{align}
It follows from the cross product that the magnetic field has an $x$- and a $z$-component, i.e. $\vec{h}= h_x\hat{x} + h_z\hat{z}$. We obtain $h_x$ and $h_z$ by taking the inverse Fourier transforms of Eq.~(\ref{hqk}). We are interested in the magnetic field at the surface of the superconducting, which is why we take $z=0$. 

The $x$-component gives an improper integral over $k$, which Cauchy principal value is equal to zero. Therefore, $h_x^\text{soc}=0$. The $z$-component is given by
\begin{equation*}
    h^\text{soc}_z = \frac{i\alpha\phi_0}{2\pi^3}\int \frac{\delta(q_y)}{|\vec{q}|^2+k^2} \frac{|\vec{q}|\sin\left(\tfrac{1}{2}q_xL\right)}{1 + 2|\vec{q}|\lambda_\text{eff}}  e^{-i\vec{q}\cdot\vec{r}}d^2\vec{q}\ dk. 
\end{equation*}
The integral over $q_y$ vanishes. The integral over $k$ yields $\pi/|q_x|$. What remains is
\begin{align*}
    h^\text{soc}_z &= \frac{i\alpha\phi_0}{2\pi^2}\int_{-\infty}^\infty\frac{\sin\left(\tfrac{1}{2}q_xL\right)e^{-iq_xx}}{1 + 2|q_x|\lambda_\text{eff}} dq_x.
\end{align*}
We split the remaining integral in $(-\infty,0)$ and $(0,\infty)$ to take the absolute value into account. Rewriting the result gives
\small
\begin{equation}
    \vec{h}^\text{soc}(x)
    = \frac{\alpha\phi_0 \hat{z}}{2\pi^2}\int_0^\infty\frac{\cos\left(q_x(\frac{L}{2}-x)\right)-\cos\left(q_x(\frac{L}{2}+x)\right)}{1 + 2q_x\lambda_\text{eff}}dq_x.  \label{hsos}
\end{equation}
which is the relation (8) of tha main text.
\normalsize

\subsection{Interaction energy $E^\text{int}(x)$}
The interaction energy has contributions from the magnetic field ($E^\text{int,h}$) and from the combined superconductivity and SOC term as described in Eq.~(\ref{ff4}) ($E^\text{int,s}$). 
Since the magnetic field gives a volume integral, while the other term is described by a surface integral, we evaluate them separately and take the sum in the end.

We first consider the magnetic field term. We use Eq.~(\ref{fourierh}) to find the Fourier transform of $h^2$, which is given by 
\begin{align*}
    h_{q,k}^2 &= \vec{h}_{q,k} \cdot\vec{h}_{-q,k} \\
    &\sim \left[(\vec{q}+\vec{k})\times (\vec{\Phi}_q + \vec{\alpha}_q)\right]\cdot \left[(\vec{q}+\vec{k})\times (\vec{\Phi}^*_q + \vec{\alpha}^*_q)\right]
\end{align*}
where we used that $\vec{\Phi}_{-q}=\vec{\Phi}_q^*$ and $\vec{\alpha}_{-q}=\vec{\alpha}_q^*$. We apply the vector identity $(a\times b)\cdot(c\times d) = (a\cdot c)(b\cdot d)-(a\cdot d)(b\cdot c)$. Since the vortices and SOC are both located in-plane, we have
$\vec{k} \cdot \vec{\Phi}_q = \vec{k} \cdot \vec{\Phi}_q^* = \vec{k} \cdot \vec{\alpha}_q = \vec{k} \cdot \vec{\alpha}_q^* =0$.
The $\vec{q}\cdot \vec{\Phi}_q$ term turns out to be zero as well. Since $\vec{q}\perp\vec{k}$, we reduce $(\vec{q}+\vec{k})^2=\vec{q}^2+\vec{k}^2$. The remaining terms in $h_{q,k}^2$ are
\begin{align*}
    h_{q,k}^2 \sim  |\vec{\Phi}_q|^2 + \vec{\alpha}_q\cdot \vec{\Phi}^*_q+ \vec{\alpha}^*_q\cdot\vec{\Phi}_q + |\vec{\alpha}_q|^2 - (\vec{q}\cdot\vec{\alpha}_q)(\vec{q}\cdot\vec{\alpha}^*_q). 
\end{align*}
The interaction between the vortices and SOC is given by the two mixed terms $\vec{\alpha}_q\cdot \vec{\Phi}^*_q+ \vec{\alpha}^*_q\cdot\vec{\Phi}_q$.
The contribution of the interaction to the energy, originating from the magnetic field (including pre-factors) is given by
\small
\begin{align*}
E_{q,k}^\text{int,h}=
    \frac{1}{8\pi}  \left(\frac{2|\vec{q}|}{1+2|\vec{q}|\lambda_\text{eff}}\right)^2\frac{1}{|\vec{q}|^2+k^2} \left( \frac{\vec{\alpha}_q}{2e}\cdot \vec{\Phi}^*_q+ \frac{\vec{\alpha}^*_q}{2e}\cdot\vec{\Phi}_q \right).
\end{align*}
\normalsize
We substitute $\vec{\alpha}_q$ and $\vec{\Phi}_q$ from Eqs.~(\ref{fourieralpha}) and (\ref{fourierphi}).
We take the inverse Fourier transform and set $e^{iq_xx}=e^{iq_yy}=e^{ikz}=1$. We obtain
\begin{align*}
E^\text{int,h} &= -\frac{\alpha\phi_0^2}{4\pi^4}\int \frac{\delta(q_y)}{|q_x|^2+k^2} \frac{\sin^2\left(\tfrac{1}{2}q_xL\right)}{(1+2|q_x|\lambda_\text{eff})^2}   d^2\vec{q}\ dk. 
\end{align*}
The integral over $q_y$ vanishes and the integral over $k$ gives $\pi/|q_x|$, and we are left with
\begin{align}
E^\text{int,h} = -\frac{\alpha\phi_0^2}{4\pi^3}\int_{-\infty}^\infty \frac{\sin^2\left(\tfrac{1}{2}q_xL\right)}{|q_x|(1+2|q_x|\lambda_\text{eff})^2} dq_x. \label{Eint1}
\end{align}
To write the $F_\text{s}+F_\text{soc}$ term from Eq.~(\ref{ff4}) in the same form, we substitute the expression for $\vec{A}_q$ from Eq.~(\ref{Aqs}) in the integrant, such that
\begin{equation*}
    \left( \vec{\Phi} -\vec{A} + \frac{\vec{\alpha}\phi_0}{2\pi} \right)^2 = \left[ \frac{2|\vec{q}|\lambda_\text{eff}}{1+2|\vec{q}|\lambda_\text{eff}} \left(\vec{\Phi}_q + \frac{\vec{\alpha}_q\phi_0}{2\pi} \right)\right]^2.
\end{equation*}
We work out the brackets and only keep the mixed terms $\vec{\alpha}_q\cdot \vec{\Phi}^*_q+ \vec{\alpha}^*_q\cdot\vec{\Phi}_q$. The corresponding interaction energy is given by
\begin{align}
    E^\text{int,s} = -\frac{\alpha\phi_0^2\lambda_\text{eff}}{2\pi^3} \int_{-\infty}^\infty \frac{\sin^2\left(\tfrac{1}{2}q_xL\right)}{(1+2|q_x|\lambda_\text{eff})^2} dq_x. \label{Eint2}
\end{align}
We are interested in the total interaction energy $E^\text{int}$, which is given by the sum of Eqs.~(\ref{Eint1}) and (\ref{Eint2}). Again splitting the integral into $(-\infty,0)$ and $(0,\infty)$, we finally obtain the total interaction energy
\begin{equation}
    E^\text{int} = -\frac{\alpha\phi_0^2}{2\pi^3}\int_0^\infty \frac{\sin^2\left(\tfrac{1}{2}q_xL\right)}{q_x(1+2q_x\lambda_\text{eff})} dq_x. \label{E}
\end{equation}
This interaction energy only describes the interaction with a vortex at $x=L/2$.
To describe the interaction with a vortex at a general position $x=x_0$, we translate the Fourier transform of the vortex as follows:
\begin{equation*}
    \vec{\Phi}_q(x_0) = e^{iq_xx_0}\vec{\Phi}_q(0)
    = e^{iq_x(x_0-L/2)}\vec{\Phi}_q(\tfrac{1}{2}L).
\end{equation*}
We repeat the calculation for this general $x_0$. Since this holds for all $x_0\in\mathbb{R}$, we may replace $x_0$ by $x$. We can now express the interaction energy as a function of $x$ as
\begin{equation}
    E^\text{int}(x) = -\frac{\alpha\phi_0^2}{2\pi^3}\int_0^\infty \frac{\sin\left(\tfrac{1}{2}q_xL\right)\sin\left(q_x x\right)}{q_x(1+2q_x\lambda_\text{eff})} dq_x. \label{Eints}
\end{equation}

\subsection{Thick ferromagnet}
To derive a criterion for vortex generation, we first need to estimate the interaction energy~(\ref{E}) in the case of a thick ferromagnet, i.e. $L\gg 1$. We introduce the substitutions
$y=2q\lambda_\text{eff}$, $p = L/4\lambda_\text{eff}$ and $C = -\alpha\phi_0^2/2\pi^3$.
Since $L\gg 1$, $p\gg 1$. We split the integral range in $[0,1/p)$ and $[1/p,\infty)$, that is
\begin{align}
    E^\text{int} = C\int_0^{1/p} \frac{\sin^2(py)}{y(1+y)}dy + C\int_{1/p}^\infty \frac{\sin^2(py)}{y(1+y)}dy.
    \label{Ethick}
\end{align}
Since $p\gg 1$, the range $[0,1/p)$ is very small. On this small interval, $\sin^2(py)$ is bounded by $p^2y^2$. Since $y\ll 1$, we estimate the denominator by $1+y\approx 1$. We obtain
\begin{align*}
C\int_0^{1/p} \frac{\sin^2(py)}{y(1+y)}dy &< Cp^2\int_0^{1/p} \frac{y}{1+y}dy \nonumber \\ &\approx Cp^2\int_0^{1/p} y \ dy = \frac{C}{2p}.
\end{align*}
For the second integral in Eq.~(\ref{Ethick}), we use $\sin^2(py)=\tfrac{1}{2}(1-\cos (2py))$, which yields again two integrals. The first one can be estimated as
\begin{align*}
 \int_{1/p}^\infty \frac{1}{y(1+y)}dy = \ln(y)-\ln(1+y)\Big|_{1/p}^\infty\approx \ln(p).
\end{align*}
To evaluate the integral with $\cos(2py)$, we introduce another substitution $u=2py$. Carefully taking the integration limits into account, we get
\begin{align*}
    \int_{1/p}^\infty \frac{\cos(2py)}{y(1+y)}dy = \int_2^\infty \frac{\cos u}{u}du + \int_2^\infty \frac{\cos u}{u+2p}dy.
\end{align*}
For $p\gg 1$, the first term is just a constant smaller than 1, while the second term vanishes. To summarise, the first integral in Eq.~(\ref{Ethick}) follows $\sim 1/p$, whereas the second integral $\sim \ln(p)$. Hence, for $p\gg 1$, the interaction energy for thick ferromagnets can be estimated by
\begin{equation*}
  E^\text{int} \simeq -\frac{\alpha\phi_0^2}{4\pi^3}\ln\left(\frac{L}{4\lambda_\text{eff}}\right).
\end{equation*}
which is the relation (9) of the main text.

\subsection{Ginzburg-Landau equation}
We consider the free energy $F=\int f\ d^2\vec{r}$ where $f$ is given by Eq.~(1) in the main text. For simplicity, we neglect the magnetic terms $\vec{A}$ and $\vec{h}$. Since the SOC is directed along the interface, i.e. $\vec{\alpha}\parallel\hat{y}$, Eq.~(1) in the main text simplifies to
\begin{align*}
    f = & a|\psi|^2 + \frac{b}{2}|\psi|^4  +\frac{1}{4m}\left( \left|\frac{\partial \psi}{\partial x} \right|^2 + \left|\frac{\partial \psi}{\partial y} \right|^2\right) \nonumber \\
    &+ \frac{i\alpha}{4m}\left(\psi\frac{\partial\psi^*}{\partial y} - \psi^* \frac{\partial\psi}{\partial y}\right).
\end{align*}
The Ginzburg-Landau (GL) equation is obtained by minimising the free energy $F$ using the Euler-Lagrange method. We use that $|\psi|^2=\psi\psi^*=\psi^*\psi$. Minimising with respect to $\psi^*$, all terms containing $\psi^*$ disappear. What remains is
\begin{equation}
    a\psi + b |\psi|^2\psi - \frac{1}{4m}\left(\frac{\partial^2\psi}{\partial x^2}+\frac{\partial^2\psi}{\partial y^2}\right) - \frac{i \alpha}{2m} \frac{\partial\psi}{\partial y} = 0. \label{GLs}
\end{equation}
which is the relation (13) of the main text.
Minimising $F$ with respect to $\psi$ results in the complex conjugate of Eq.~(\ref{GLs}).

\subsection{Wave function $\psi(x)$ for $T=T_c$}
At $T_c$, we have $|\psi|^2\to 0$ and the GL equation~(\ref{GLs}) is linearised. We use $\psi(x,y)=\psi(x)e^{iq_yy}$ and introduce the substitutions $\tau = 4ma/\alpha^2$, $\kappa = q_y/\alpha$, $x\mapsto \alpha x$, $L\mapsto \alpha L$. The linearised GL equation becomes
\begin{equation*}
    \begin{cases}
    \left( \tau +(\kappa+1)^2-1\right)\psi(x) - \dfrac{d^2\psi}{dx^2} = 0 &\mbox{ for }|x|<\tfrac{1}{2}L, \\
    \left( \tau +\kappa^2\right)\psi(x) - \dfrac{d^2\psi}{dx^2} = 0 &\mbox{ for }|x|>\tfrac{1}{2}L. 
    \end{cases}. 
\end{equation*}
The wave function satisfying the boundary conditions $\psi(x)=\psi(-x)$ and $\lim_{x\to \pm\infty}\psi(x)=0$ is
\begin{equation}
    \psi(x)=\begin{cases}
    \psi_1 \cos\left(\sqrt{1-\tau-(\kappa+1)^2}x\right) &\mbox{ for }|x|<\tfrac{1}{2}L, \\
    \psi_2 \exp\left(-\sqrt{\tau+\kappa^2}x\right) &\mbox{ for }|x|>\tfrac{1}{2}L. 
    \end{cases} \label{newwave}
\end{equation}
We define $\kappa=-1+\varepsilon$, where $\varepsilon$ is a small difference between the modulation vector and $\alpha$. We impose continuity of $\psi(x)$ and $\psi'(x)$ at $x=\pm L/2$, which results in a relation between $\tau$ and $\varepsilon$. Maximising $\varepsilon$ with respect to $\tau$ (i.e. $d\tau/d\varepsilon=0$), we find $\tau \approx 1- (\pi/L)^2$ and $\varepsilon \approx \pi^2/(\sqrt{2}L^3+\pi^2)$. We express $\psi_2$ in terms of $\psi_1$ and set $\psi_1=1$. The resulting wave function is shown in Fig.~\ref{Fig2}(a) in the main text.

\subsection{Current density $\vec{j}(x)$ for $T=T_c$}
The current density is obtained from $\vec{j}=-\partial f/\partial \vec{A}$. After minimising $f$, we set $\vec{A}=0$ and obtain
\begin{align}
    \vec{j} &= \frac{ie}{2m}\left[\psi^*\vec{\nabla}\psi-\psi\vec{\nabla}\psi^*\right] - \frac{e\vec{\alpha}}{m} |\psi|^2. \label{cur}
\end{align}
We substitute $\psi(x,y)=\psi(x) e^{iq_yy}$, with $\psi(x)$ given by Eq.~(\ref{newwave}) and $q_y = \kappa\alpha = \alpha(-1+\varepsilon)$.
Since $\psi(x)\in\mathbb{R}$, the component $j_x = 0$.

For $|x|>L/2$, there is no SOC, which implies the last term in Eq.~(\ref{cur}) vanishes and $\varepsilon=0$. For $|x|<L/2$, we have $q_y=\alpha(-1+\varepsilon)$, which yields a current density proportional to $\varepsilon$. The resulting current density is given by Eq.~(\ref{current}).

\onecolumngrid
\section{Ferromagnetic half plane}
\twocolumngrid

\subsection{Modulated critical temperature}
We use the GL equation~(\ref{GLs}) to investigate the transition temperature $T_c$. We first assume superconductivity is uniform in the $x$-direction, and a plane wave in the $y$-direction, i.e. $\psi(x,y) = \psi_0e^{iq_yy}$. 
Under this assumption, we may linearise the GL equation~(\ref{GLs}), i.e. neglect the $b |\psi|^2\psi$ term. Substituting $\psi(x,y)$ into the linearised GL equation, we obtain a quadratic equation in $q_y$. Taking the derivative and setting it equal to zero, we find that $T-T_c$ has a maximum for $q_0\equiv-\alpha$.

We use the obtained value for $q_0$ to study the more general case, in which we no longer assume uniformity of the wave function.
Instead, we use $\psi(x,y) = \psi(x) e^{-i\alpha y}$, where we are interested in finding $\psi(x)$.
We apply separation of variables to the GL equation~(\ref{GLs}) and evaluate the derivatives with respect to $y$ separately.
We  distinguish between the regions with and without SOC. In the presence of SOC ($\alpha>0$), the $y$-dependent terms in the Ginzburg-Landau equation~(\ref{GLs}) become
\begin{equation*}
    -\frac{1}{4m}\frac{\partial^2\psi}{\partial y^2} - \frac{i \alpha}{2m} \frac{\partial\psi}{\partial y} 
    = - \frac{\alpha^2}{4m}.
\end{equation*}
In the absence of SOC ($\alpha=0$), we only have the $\partial^2/\partial y^2$ term, which gives a contribution of $\alpha^2/4m$.

For a fixed value of $\alpha$, these terms are just a constant offset to the critical temperature. We define the new modified critical temperature in the presence of SOC as
\begin{align}
    T_{c+} = T_c \left( 1 + \frac{\alpha^2}{4ma_0}\right). \label{Tcs}
\end{align}
which corresponds to the formula (15) of the main text. Note that the T$_c$ modulation in the formula (\ref{Tcs}) is the same for the case of the ferromagnetic strip.

\subsection{Wave function $\psi(x)$}
We calculate the wave function $\psi(x)$ in the regions $x>0$ and $x<0$ separately and apply appropriate boundary conditions at $x=0$. 
The GL equation~(\ref{GLs}) for $x>0$ ($x<0$) can be written as 
\begin{equation*}
  a_0 \frac{T-T_{c+}}{T_c}\psi + b |\psi|^2\psi - \frac{1}{4m}\frac{d^2\psi}{dx^2} = 0. 
\end{equation*}
For $x\gg 0$, we neglect the second derivative with respect to $x$ and obtain
\begin{equation*}
   |\psi|^2 = -\frac{a_0}{b} \frac{T-T_{c+}}{T_c}.
\end{equation*}
We note that $|\psi|^2>0$. For convenience of writing, we define $\psi_\infty \equiv \sqrt{|\psi|^2}$ and write the wave function as $\psi(x) = \psi_\infty f(x)$, where $0\le f(x)\le 1$ takes the variation in space into account.
We define the temperature dependent coherence length $\xi(T)$ as
\begin{equation*}
    \xi^2(T) = \frac{1}{4m}\frac{T_c}{a_0|T-T_{c+}|}.
\end{equation*}
For $x>0$, we obtain the GL equation in terms of $f$:
\begin{equation}
    - \xi^2(T)\dfrac{d^2f}{dx^2} - f + f^3 = 0. \label{GLnew}
\end{equation}
We multiply Eq.~(\ref{GLnew}) by $df/dx$ and rewrite it as
\begin{equation*}
  - \frac{\xi^2(T)}{2}\frac{d}{dx}\left(\frac{df}{dx}\right)^2- \frac{d}{dx}\left(\frac{f^2}{2}\right) + \frac{d}{dx}\left(\frac{f^4}{4}\right) = 0.
\end{equation*}
We integrate this equation. Since $\lim_{x\to\infty} f(x) = 1$, the integration constant is equal to $-\tfrac{1}{2}$ and the Eq.~(\ref{GLnew}) becomes
\begin{equation*}
   \xi^2(T)\left(\frac{df}{dx}\right)^2 = \tfrac{1}{2}(1-f^2)^2. \label{eqf}
\end{equation*}
We can follow the same steps for $x<0$. The boundary condition $\lim_{x\to -\infty}f(x)=0$ results in the integration constant being equal to 0. Using the definitions of $\psi_\infty$ and $\xi(T)$, we obtain
\begin{align*}
     & \xi^2(T)\left(\frac{df}{d x}\right)^2 = \tau f^2 + \tfrac{1}{2} f^4,
    && \tau \equiv \frac{T-T_{c}}{|T-T_{c+}|}.
\end{align*}
We these differential equations for $f(x)$ and use that the wave function is conditions at $x=0$ and equal to $f(0)\equiv f_0$, for some $0<f_0<1$. Substituting $f(x)$ back into $\psi(x)=\psi_\infty f(x)$, we find
\begin{equation*}
    \psi(x) = \psi_\infty\begin{cases}
    -\dfrac{4\tau c_2 e^{\sqrt{\tau}x/\xi}}{2\tau c_2^2e^{2\sqrt{\tau}x/\xi}-1} & \mbox{ for }x<0,\\\\
    \tanh\left(\dfrac{x}{\sqrt{2}\xi}+c_1\right) & \mbox{ for }x>0,
    \end{cases}
\end{equation*}
with 
\begin{align*}
    & c_1 = \tfrac{1}{2}\ln\left(\frac{1+f_0}{1-f_0}\right),
    && c_2 = \frac{-2\tau+\sqrt{4\tau^2+2\tau f_0^2}}{2\tau f_0},
\end{align*}
where we obtain $f_0$ from continuity of the first derivative at $x=0$; $f_0 = 1/\sqrt{2(\tau+1)}$.

\subsection{Vector potential $\vec{A}(x)$}
Applying Biot-Savart's law $\vec{A} = \int \vec{j}/\sqrt{x^2+y^2} \ d\vec{l}$ to the current (relation (16) in the main text) results in an implicit equation for $A_y(x,z)$.
We introduce the dimensionless lengths $x \mapsto x/\lambda_\text{eff}$, $y \mapsto y/\lambda_\text{eff}$ to obtain a dimensionless form. The implicit equation for $A_y(x,z)$ in the $z=0$ plane is given by
\begin{align*}
    A_y(x,z=0) &= \int_{-\infty}^\infty \frac{j_0}{\sqrt{x^2+y^2}}\ dy \\
    &- \frac{1}{4\pi}\int_{-\infty}^\infty \int_0^\infty \frac{A_y(x')}{\sqrt{(x-x')^2+y^2}}\ dx' \ dy.
\end{align*}
This expression is for a ferromagnet covering the half-plane, i.e. $x\in[0,\infty)$. To avoid possible problems with convergence, we consider a ferromagnetic strip instead, i.e. $x\in[0,L]$. This means that at the other end of the strip we have to add a current $-j_0\delta(x-L)\hat{y}$. The integration range for $x'$ is now $[0,L]$. Since these integrals are even in $y$, we use that the integral over $(-\infty,\infty)$ is equal to twice the integral over $[0,\infty)$. The expression for $A_y(x)$ becomes
\begin{align*}
    A_y(x) = &\ 2j_0\int_0^\infty \frac{1}{\sqrt{x^2+y^2}}-\frac{1}{\sqrt{(x-L)^2+y^2}}\ dy \\
    &- \frac{1}{2\pi}\int_0^\infty \int_0^L \frac{A_y(x')}{\sqrt{(x-x')^2+y^2}} dx' \ dy.
\end{align*}
To remove the prefactor, we make the substitution $A_y\mapsto A_y/2j_0$.
Performing integration over $y$, we obtain the expression for $A_y(x)$ in Eq.~(17) of the main text.